# Terahertz two-dimensional coherent spectroscopy: unlocking coherent quantum intraband nonlinear dynamics

Se Jin Park[1], In Hyeok Choi[2], and Jeong Woo Han[1,3*]

[1]Department of Physics Education and Center for Quantum Technologies, Chonnam National University, Gwangju 61186, South Korea

[2]Department of Chemistry, Massachusetts Institute of Technology, Cambridge, Massachusetts 02138, USA

[3]Graduate School of Interdisciplinary Program for Photonic Engineering, Chonnam National University, Gwangju 61186, South Korea

*Corresponding author: jwhan@chonnam.edu

Quantum materials host a rich spectrum of low-energy collective excitations- phonons, magnons, plasmons, polaritons, and Higgs modes- that fundamentally govern their emergent properties, but remain inaccessible to conventional linear THz spectroscopy when nonlinear pathways overlap or non-perturbative dynamics emerge. Terahertz two-dimensional coherent spectroscopy (THz-2DCS), realized through phase-stable high-field THz pulse sequences, directly resolves this challenge by delivering phase-resolved multidimensional spectra that uniquely disentangle coherent intraband nonlinear quantum dynamics from incoherent backgrounds. Here, we establish a unified theoretical framework by using density-matrix formalism combined with the optical Bloch equations, from which nonlinear pathways are systematically derived and visualized on the Bloch sphere. This framework provides insight into low-energy nonlinear dynamics exhibiting homogeneous or inhomogeneous broadening depending on the scattering mechanism, enabling the interpretation of nonlinear signals in both the perturbative regime, described by a susceptibility expansion, and the non-perturbative regime. Building on this foundation, we comprehensively survey THz-2DCS investigations of nonlinear plasmon, phonon, magnon, and polariton. Finally, we outline future opportunities, including integration with optical-pump THz-probe platforms for nonequilibrium quantum control.

## 1. Introduction

Quantum materials exhibit a variety of emergent phenomena arising from strongly coupled collective degrees of freedom, including lattice vibrations (phonons), spin excitations (magnons), charge fluctuations (plasmons), superconducting order parameters (Higgs modes), and other low-energy quasiparticles [1-9]. Most of these elementary excitations reside in the terahertz (THz) frequency range and govern the fundamental optical, magnetic, and electronic properties of quantum materials. Although conventional linear THz spectroscopic techniques, e.g., THz-time-domain spectroscopy (THz-TDS), have provided invaluable insights into these low-energy excitations [10,11], they often fail to disentangle multiple nonlinear processes when different collective modes overlap within the same spectral window. As a result, there has been a growing demand for multidimensional spectroscopic techniques capable of resolving the microscopic origin of complex many-body interactions while preserving the coherent information encoded in quantum states [12-14].

One-dimensional nuclear magnetic resonance spectroscopy (1D NMR), employing a radio-frequency (RF) pulse as a probe beam, offers rich information on the chemical environment and molecular structure [15]. However, when a target sample is a complex compound system, such as biomolecules, the resonance modes for individual constituent ions can appear within the same frequency range, hindering clarification of the origin of each resonance. To disentangle these overlapped signals, 1D NMR was extended into two-dimensional NMR spectroscopy (2D NMR) by incorporating an additional pulse sequence that serves as a pump pulse [16]. This approach decodes the signals by tracking the phase evolution of quantum states during the time delay between pulses, thereby visualizing light–matter interactions as a coherent two-dimensional map. Richard R. Ernst was awarded the 1991 Nobel Prize in Chemistry for his pioneering work in the development of 2D NMR. His work inspired the extension of multidimensional coherent spectroscopy into infrared and optical frequency regimes. However, because conventional optical detectors measure only light intensity, extending multidimensional spectroscopy into higher-frequency optical regimes inevitably results in the loss of phase information, requiring sophisticated interferometric techniques, such as optical heterodyne detection, to reconstruct the optical phase.

Besides multidimensionality, another important aspect of spectroscopic techniques is the accessible frequency window, which determines the researchable physics. This has prompted scientific endeavors to expand the frequency range. For example, the lower frequency limit of conventional Fourier-transform infrared spectroscopy (FT-IR) is typically restricted to the far-infrared region, corresponding to wavelengths of approximately 100 μm [17]. Below this limit lies the THz frequency range, where a wide variety of collective excitations, including phonons, magnons, plasmons, and Higgs mode, can be effectively investigated [18-20]. Furthermore, the characteristic relaxation times of these quasiparticles are typically on the picosecond timescale, matching the temporal duration of THz electromagnetic pulses. As a result, there has been a strong demand to extend spectroscopic techniques beyond the frequency limit of FT-IR into the THz regime, leading to the development of THz-TDS.

Unlike conventional optical spectroscopies that are performed in the frequency domain, THz-TDS directly measures the electric field of the THz pulse in the time domain. By Fourier transforming the measured waveform, both the amplitude and phase of the THz electric field are independently obtained without requiring phase reconstruction. This means that two independent, reliable pieces of information, amplitude and phase, can be obtained directly from experimental THz-TDS data recorded in the time domain, allowing these two quantities to be directly converted into complex optical response functions (e.g., complex optical conductivity) consisting of a real and an imaginary part [21,22]. More importantly, this phase-resolved nature enables precise synchronization between THz electromagnetic fields and quantum states of matter because of the low-energy nature of the THz light, providing a natural platform for coherently manipulating low-energy collective excitations.

While multidimensional spectroscopy extends the capability of disentangling overlapping quantum pathways, it does not by itself broaden the range of accessible physical phenomena beyond linear responses. To broaden the boundaries of investigable physics, thereby approaching nonlinear phenomena, an intense light source is inevitably required. The rapid development of femtosecond ultrafast laser technology has enabled the generation of high-field THz pulses through nonlinear optical processes, opening a new avenue for nonlinear THz spectroscopy [23-25]. Upon increasing optical power further, the THz fields directly drive stronger nonlinear collective motions of quasiparticles, allowing access to the emergence of new physical phenomena, non-perturbative region, which are not described by the expansion of electrical susceptibility (perturbative region) [26-29].

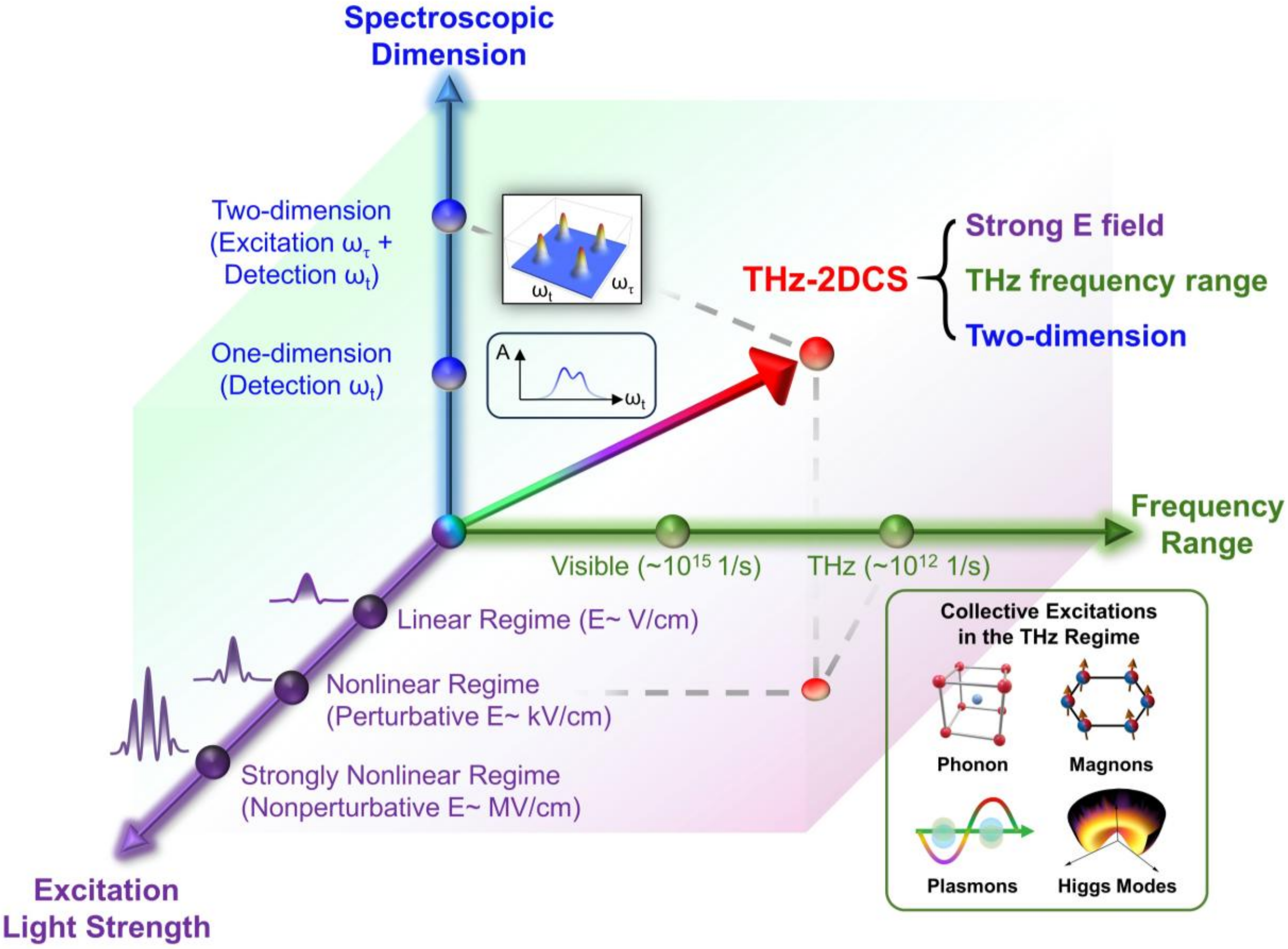

**FIGURE 1 |** Schematic diagram illustrating the development of optical spectroscopies. The advancement of spectroscopy has progressed toward expanding its frequency range, spectroscopic dimension, and increasing the power of the incident optical pulses. By approaching the THz frequency range and expanding spectroscopic dimensions, coherent quantum dynamics of low-energy collective excitations, such as phonons and magnons, can be investigated. Furthermore, increasing the strength of the illuminated electric field drives the physical regime into nonlinear domains, and even into a non-perturbative regime that cannot be described by standard mathematical expansion. Terahertz two-dimensional coherent spectroscopy (THz-2DCS) represents one of the advanced spectroscopic techniques realized by pushing the boundaries across these three metrics.

We therefore introduced three major milestones in the development of modern spectroscopy: (1) the extension of the spectral windows into the THz frequency region, (2) the securing of high-field light sources enabled by ultrafast laser technology for the investigation of nonlinear phenomena, and (3) the expansion into multidimensional spectroscopy pioneered by NMR. The convergence of these three advances has led to the development of THz two-dimensional coherent spectroscopy (THz-2DCS), in which high-field THz pulses are able to coherently excite nonlinear collective excitations in quantum materials. Because THz-2DCS naturally preserves both amplitude and phase information of the emitted THz field, it provides phase-resolved multidimensional spectra capable of disentangling overlapping nonlinear pathways and distinguishing coherent quantum dynamics from incoherent processes such as laser-induced heating or population relaxation. As a result, THz-2DCS has rapidly emerged as one of the most powerful spectroscopic techniques for investigating nonequilibrium nonlinear dynamics and low-energy many-body interactions in quantum materials. The schematic illustration for the development direction of optical spectroscopies is summarized in Fig. 1.

This review article provides a systematic and comprehensive overview of terahertz two-dimensional coherent spectroscopy (THz-2DCS), focusing on both its fundamental physical principles and the analytical framework required to interpret multidimensional THz spectra. We begin by introducing the theoretical foundations of nonlinear and quantum optics that describe the interaction between phase-locked THz pulse sequences and quantum systems. Building upon this framework, we discuss the experimental implementation and analytical methodologies used to decode multidimensional spectra and identify distinct nonlinear quantum pathways. Finally, we review representative studies of THz-2DCS to a broad range of collective excitations, including phonons, magnons, plasmons, and polaritons, and conclude by discussing the current challenges and future opportunities of THz-2DCS as a versatile platform for probing and coherently controlling quantum materials.

## 2. Theoretical background

As suggested by its name, THz-2DCS exploits two phase-locked high-intensity THz pulses that interact with a target sample as the pump and the probe beams. These high field THz pulses can be generated by non-centrosymmetric inorganic crystals such as ZnTe [30,31], GaP [32], GaSe [33,34], and $LiNbO_3$, or organic crystals like DAST and DSTMS [35-38]. Figure 2(a) shows two THz pulses, denoted by $\mathbf{E}_A$ and $\mathbf{E}_B$, incident on the sample with a delay time $\tau$, where $\mathbf{E}_A$ precedes $\mathbf{E}_B$. Thus, it is essential to understand nonlinear light-matter interactions, especially second- and third- order nonlinear processes, which correspond to two interactions (one for each pulse) and three interactions in total (two for one pulse, either $\mathbf{E}_A$ or $\mathbf{E}_B$, and one for the remaining pulse), respectively. From now on, we will discuss the detailed background physics for understanding these nonlinear interactions.

## 2.1. Nonlinear Optical Response (perturbative region)

In the perturbative regime, the polarization of a medium **P**(**E**) is induced by an incident THz electric field **E**(t) and it can be expanded by the polynomial form as [39-41]:

$$\mathrm{P}(\mathbf{E})=\varepsilon_0\left(\chi^{(1)}|\mathbf{E}|+\chi^{(2)}|\mathbf{E}|^2+\chi^{(3)}|\mathbf{E}|^{3\cdots}\right) \tag{1}$$

where $\varepsilon_0$ is the vacuum permittivity and the $n$ th order nonlinear susceptibility tensor is denoted by $\chi^{(n)}$. To quantitatively understand the nonlinear light-matter interactions, the Maxwell wave equation should be considered, combined with the nonlinear polarization as the source term, as:

$$\nabla^2\mathbf{E}-\frac{n_0}{\mathrm{c}^2}\frac{\partial^2\mathbf{E}}{\partial \mathrm{t}^2}=\mu_0\frac{\partial^2\mathbf{P}_{\mathrm{NL}}}{\partial \mathrm{t}^2} \tag{2}$$

Here, $n_0$ is the linear refractive index of the medium, $c$ is the speed of light, and $\mu_0$ is the permeability of a vacuum. $\mathbf{P}_{\mathrm{NL}}$ is the nonlinear polarization, which refers to the higher-order polarizations beyond the linear term in Eq. (1). To illustrate detailed nonlinear behaviors, we first examine the second-order nonlinear optical process as a representative case. By substituting $\mathbf{E}(\mathrm{t})=\sum_{q=1}^{2}\frac{1}{2}\mathrm{E}_l\left(\omega_q\right)\mathrm{e}^{-i\omega_q\mathrm{t}}+\mathrm{c.c.}$ into $\mathbf{E}(\mathrm{t})$ in the second-order nonlinear polarization $\mathbf{P}_i^{(2)}(\mathrm{t})=\varepsilon_0\chi_{ijk}^{(2)}\mathbf{E}_j(\mathrm{t})\mathbf{E}_k(\mathrm{t})$, four nonlinear interactions, serving as the source term in Eq. (2), can be obtained, as:

$$\mathbf{P}_i^{(2)}\left(2\omega_q\right)=\frac{1}{2}\varepsilon_0\,\chi_{ijk}^{(2)}\mathbf{E}_j\left(\omega_q\right)\mathbf{E}_k(\omega_q) \tag{3}$$

$$\mathbf{P}_i^{(2)}(\omega_1+\omega_2)=\varepsilon_0\chi_{ijk}^{(2)}\mathbf{E}_j(\omega_1)\mathbf{E}_k(\omega_2) \tag{4}$$

$$\mathbf{P}_i^{(2)}(\omega_1-\omega_2)=\varepsilon_0\chi_{ijk}^{(2)}\mathbf{E}_j(\omega_1)\mathbf{E}_k^*(\omega_2) \tag{5}$$

$$\mathbf{P}_i^{(2)}(0)=\frac{1}{2}\varepsilon_0\chi_{ijk}^{(2)}[\mathbf{E}_j(\omega_1)\mathbf{E}_k^*(\omega_1)+\mathbf{E}_j(\omega_2)\mathbf{E}_k^*(\omega_2)] \tag{6}$$

Here, $i$,$j$,$k$ represent the Cartesian spatial components, and c.c. is the corresponding complex conjugate, denoted by an asterisk. We considered the two monochromatic waves indicated by the subscript $q$. Equations (3) to (6) are responsible for the second-harmonic generation, sum-frequency generation, difference-frequency generation, and optical rectification, respectively, hereafter acronym as SHG, SFG, DFG, and OR. From them, one can find an important rule that the complex conjugate of $\mathbf{E}$(t) serves as a result of subtracting the oscillating frequency for the induced polarization.

While second-order processes require non-centrosymmetric media, third-order nonlinear effects can occur in all materials regardless of the existence of inversion symmetry. Thus, we can naturally extend this framework to third-order interactions. As we mentioned earlier, THz-2DCS employed two identical pulses of $\mathbf{E}_A$ and $\mathbf{E}_B$ separated with the delay time $\tau$, and THz pulses are recorded probe time t. Note that we denote the center frequency of $\mathbf{E}_A$ and $\mathbf{E}_B$ as $\omega_0$. In the same manner of the second-order nonlinearity with the condition of THz-2DCS, the possible third-order nonlinear polarization can be expanded, as [42]:

$$\begin{aligned} \mathbf{P}_i^{(3)} \propto\, & \mathbf{P}^{(3)}_{A+B-B}\mathbf{E}_A\mathbf{E}_B\mathbf{E}_B^* e^{-i\omega_0(t+\tau)} + \mathbf{P}^{(3)}_{-A+B+B}\mathbf{E}_A^*\mathbf{E}_B\mathbf{E}_B e^{-i\omega_0(t-\tau)} \\ & + \mathbf{P}^{(3)}_{-A+A+B}\mathbf{E}_A^*\mathbf{E}_A\mathbf{E}_B e^{-i\omega_0(t)} + \mathbf{P}^{(3)}_{A+A-B}\mathbf{E}_A\mathbf{E}_A\mathbf{E}_B^* e^{-i\omega_0(t+2\tau)} \\ & + \mathbf{P}^{(3)}_{A+B+B}\mathbf{E}_A\mathbf{E}_B\mathbf{E}_B e^{-i\omega_0(3t+\tau)} + \mathbf{P}^{(3)}_{A+A+B}\mathbf{E}_A\mathbf{E}_A\mathbf{E}_B e^{-i\omega_0(3t+2\tau)} + \text{c.c} \end{aligned} \tag{7}$$

Here, $\mathbf{E}_A$ moves along with $\tau$, while the maximum position of $\mathbf{E}_B$ is fixed at t=0, defining $\mathbf{E}_A(t, \tau)$ and $\mathbf{E}_B(t)$ (see Fig. 2(a) and (b)). The nonlinear signals $\mathbf{E}_{NL}$ induced by joint efforts of $\mathbf{E}_A$ and $\mathbf{E}_B$ (denoted by $\mathbf{E}_{AB}$) can be isolated by subtracting the individual responses, as [43-45]:

$$\mathbf{E}_{NL}(t, \tau) = \mathbf{E}_{AB}(t, \tau) - \mathbf{E}_A(t, \tau) - \mathbf{E}_B(t) \tag{8}$$

From now on, we adopt the convention where the complex conjugate is denoted by a minus sign. The first and second terms in Eq. (7), A+B-B and -A+B+B, are known as the AB non-rephasing (NR) and AB photon echo signals (Echo), respectively, where "AB" indicates that $\mathbf{E}_A$ arrives before $\mathbf{E}_B$ [46]. Although only the AB sequence is covered here, it is emphasized that there are also the corresponding nonlinear signals in the BA sequence. Due to temporal causality, time t always flows in the positive direction. Therefore, the phase term of the AB photon echo signal can vanish at $t = \tau$, indicating that the phase recovers its initial state prior to light interaction. In contrast, the AB non-rephasing signal does not exhibit this behavior, from which the first and the second terms are named as "non-rephasing" and "echo", respectively. Importantly, both signals show a dependence on $\tau$, implying that phase-related information is engraved by the $\mathbf{E}_A$ into the system. Therefore, the AB non-rephasing and AB photon echo signals are relevant to the superposition of two quantum states (i.e., quantum coherent states). Note that NR and Echo signals can be referred to as different names of "Free-induction decay" and "Rephasing" signals, respectively.

The third term is referred to as the AB pump–probe (PP) signal, which is equivalent to the signal obtained in conventional pump–probe spectroscopy [47]. In a typical pump–probe experiment, the pump pulse generates a population inversion through the absorption process, thereby increasing the population of excited carriers, while the probe pulse monitors their subsequent dynamics, such as population relaxation and recombination. Since the absorption process is associated with population transfer rather than quantum coherence, the PP signal is independent of the phase evolution with respect to the pump delay, as indicated by the absence of the delay-dependent phase term in Eq. (7). Specifically, the two successive interactions with $\mathbf{E}_A$ (-A+A) create the population inversion, which is subsequently interrogated by the interaction with $\mathbf{E_B}$. Therefore, the preceding pulse in THz-2DCS plays exactly the same role as the pump pulse in conventional pump–probe spectroscopy.

The fourth term corresponds to the second-quantum (2Q) signal, which originates from two successive interactions with $\mathbf{E}_A$ (A+A). These interactions create a two-quantum coherence that accumulates phase at twice the rate of the fundamental coherence, resulting in a phase evolution of $2\tau$. This doubled phase accumulation gives rise to the name “second-quantum” signal. The subsequent interaction with $\mathbf{E}_B$ serves as a readout pulse that converts the 2Q into a radiative one-quantum coherence (1Q), allowing the encoded 2Q information to be detected. Since second-harmonic generation (SHG) (see Eq. (3)) also involves twice the interactions with the driving field, it is closely related to two-quantum coherence. In SHG, however, the two-quantum coherence radiates directly at twice the fundamental frequency of the incident light field, producing the second-harmonic field. As a result, efficient SHG can only be observed when the phase-matching condition is satisfied. In contrast, in THz-2DCS, the two interactions with $\mathbf{E}_A$ encode the 2Q coherence. The subsequent interaction with $\mathbf{E}_B$ converts it into the 1Q coherence that emits the same frequency as the incident light, suggesting the 2Q signal can be detected even when the direct second-harmonic emission is not phase matched [48]. It is worth noting that, in a perfectly harmonic system with equally spaced energy levels, destructive interference between pathways, specifically the cancellation of emissions from the first and second excited states, causes the 2Q signal to vanish [8]. Therefore, the observation of the finite 2Q signal generally indicates the presence of anharmonicity or many-body interactions in the system. The last two terms are associated with the four-wave mixing signals.

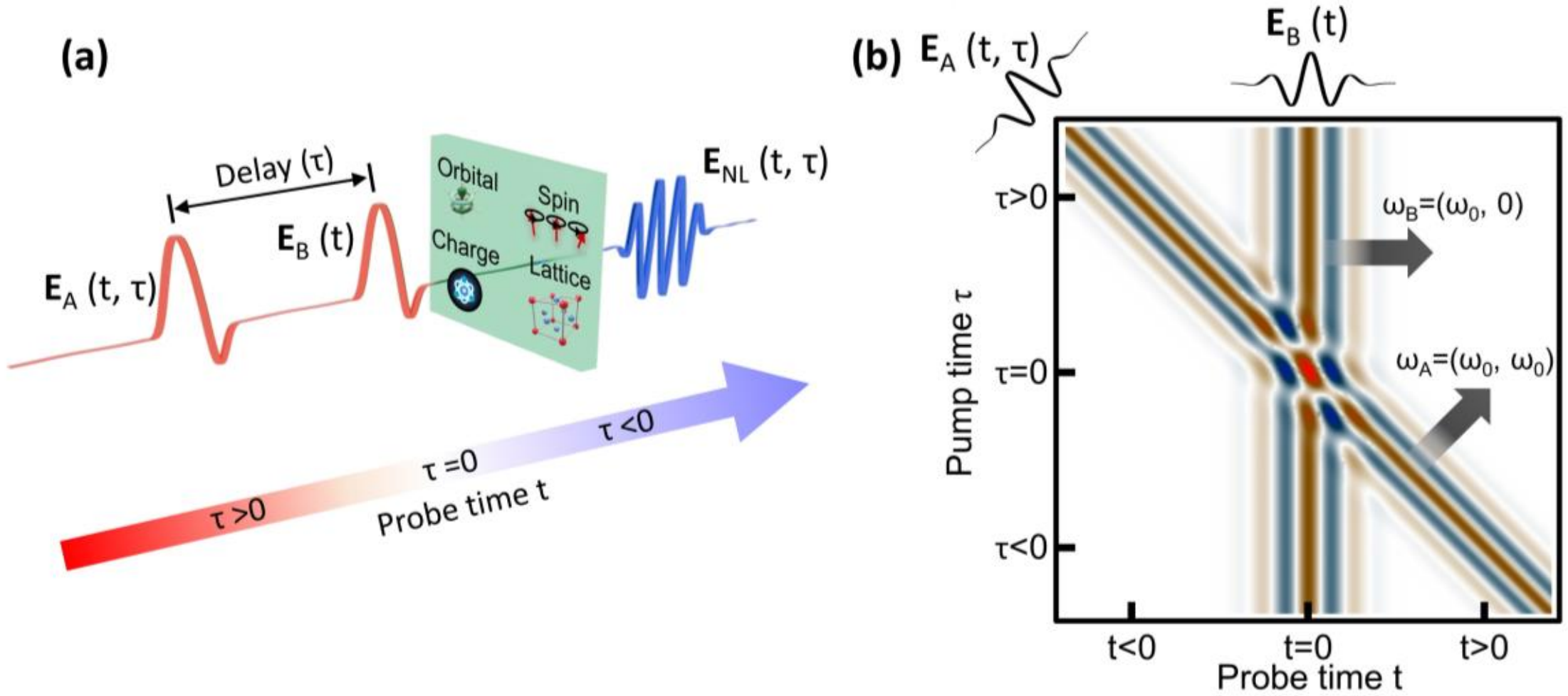

**FIGURE 2 |** Measurement scheme of terahertz-two-dimensional coherent spectroscopy (THz-2DCS). (a) Schematic illustration of THz-2DCS. Two terahertz excitation pulses, $E_A$ and $E_B$, are incident on quantum materials to excite the low-energy collective mode of the quantum degrees of freedom (orbital, spin, charge, and lattice), emitting nonlinear signals $E_{NL}$. $E_A$ and $E_B$ are temporally separated by the delay time τ and are recorded in the probe time t. The exact overlapping position between $E_A$ and $E_B$ is defined at τ=0 and $E_B$ is fixed at τ =0 such that $E_B$ is a function of t only, i.e., $E_B(t)$. Whereas the position of $E_A$ is varied, depending on t and τ, resulting in $E_A= E_A(t, \tau)$ and $E_{NL} =E_{NL}(t, \tau)$. The sign convention of τ is plus (minus) when $E_A(E_B)$ precedes $E_B(E_A)$. (b) 2D map of the incident THz electric fields of $E_A(t, \tau)$ and $E_B(t)$ as a function of the probe time t and the pump time τ. The normal vectors of the phase fronts correspond to the frequency vectors, i.e., $\omega_A=(t=\omega_0, \tau =\omega_0)$ and $\omega_B=(\omega_0, 0)$.

Figure 2(a) shows the electric field profiles of $\mathbf{E}_A$ and $\mathbf{E}_B$ as the two-dimensional map as a function of the probe time t (horizontal axis) and the pump delay τ. The phase front of $\mathbf{E}_A$ is tilted by 45 degrees, meaning $\mathbf{E}_A$ depends on both t and τ, contrasting to $\mathbf{E}_B$ that is only determined by t, leading to the horizontal phase front. Given the phase front of $\mathbf{E}_A$ and $\mathbf{E}_B$, one can obtain the corresponding frequency vectors, i.e., $\omega_A=(\omega_t=\omega_0, \omega_\tau=\omega_0)$ and $\omega_B=(\omega_0,0)$ [28,49-51]. Note that we herein defined τ>0 when $\mathbf{E}_A$ is before $\mathbf{E}_B$ and the sign of $\omega_\tau$ becomes minus if τ<0, i.e., $\omega_A=(\omega_0, -\omega_0)$, corresponding to the phase accumulation directions of $\delta(t+\tau)$ and $\delta(t-\tau)$, respectively. The sign conversion on THz-2DCS is crucial for the assignment of the nonlinear signals and is prone to a glitch, such that we need to address the wavevectors very carefully. The common measurement geometry of THz-2DCS is a collinear configuration in which $\mathbf{E}_A$ and $\mathbf{E}_B$ illuminate the target sample at normal incidence, leading to difficulty in disentangling the third-order nonlinearities, e.g., NR or Echo, in the time domain. In principle, we can obtain the individual nonlinear signals by positioning the detectors in the corresponding spatial positions; however, it is very challenging to isolate the individual signals due to the long-wavelength nature of the THz light, which accompany huge divergence properties. Since the nonlinear interactions have different Liouville space pathways that are determined by the number of interactions, and thus the pathways are represented by the respective linear combinations of the frequency vectors, it is possible to disentangle the nonlinear interactions in the frequency domain. The 2D time-domain data can be transferred to the corresponding frequency domain by a 2D Fourier transform. While these Liouville pathways are defined in the frequency domain, the equivalent time-domain representation is the Feynman double-sided diagram [52], derived from the density matrix picture of the nonlinear response [43].

### 2.2. Density Matrix Formalism

To understand the third-order nonlinear signals observable in THz-2DCS in the quantum evolution manner, a rigorous and intuitive framework for describing how the quantum state of the material evolves during successive light-matter interactions is required. The density matrix formalism offers deep insights into quantum evolution throughout the interaction sequence. This enables a clear distinction between population, related to absorption dynamics (e.g., pump-probe signals), and quantum coherence pathways (e.g., non-rephasing and photon echo signals). The density matrix **ρ** of the two-level atoms, the simplest example, is represented by a two-by-two matrix, as [53,54]:

$$\boldsymbol{\rho}=\begin{pmatrix}\rho_{00} & \rho_{10}\\ \rho_{01} & \rho_{11}\end{pmatrix} \tag{9}$$

Here, the diagonal terms ($\rho_{00}$ and $\rho_{11}$) represent the population densities of the carriers for the ground and the excited states, and the off-diagonal terms ($\rho_{10}$ and $\rho_{01}= (\rho_{10})^*$) describe the coherent superposition states. The Schrödinger picture can only represent pure states, i.e., coherent superpositions of wavefunctions with a well-defined phase, and thus it has inherent limitations in describing statistical mixtures. In other words, it is impossible to describe, for example, a mixed state in which the carriers occupy the ground and excited states with the same probability, with washed-out phase information, whereas the density matrix formalism can readily represent such a mixed state by setting the off-diagonal terms to zero, i.e., $\rho_{00}$ and $\rho_{11}$=0.5 and $\rho_{10}$ and $\rho_{01}$=0. For any physical density matrix, the off-diagonal elements should satisfy $|\rho_{10}|^2 \leq \rho_{00}\rho_{11}$, known as Cauchy–Schwarz inequality [55]. Therefore, when $\rho_{00}=\rho_{11}=0.5$, the maximum coherence is $|\rho_{10}|=0.5$, whereas a completely mixed state corresponds to $\rho_{10}=\rho_{01}=0$. This demonstrates that mixed states cannot be represented by a single wavefunction but are naturally described within the density-matrix formalism.

Here, we present the detailed quantum coherent evolution of two representative third-order nonlinear processes associated with coherent evolution and population dynamics: the photon-echo and pump-probe signals. Figure 3(a) shows the density-matrix representation of the photon-echo (echo) third-order nonlinear signal using the Bloch sphere and the double-sided Feynman diagram. In the Bloch sphere, the equatorial plane is formed by the two axes, **u** and **v**, which are defined by the real and imaginary parts of $\rho_{10}$, respectively. Since the imaginary component is 90° out of phase with the incident field, the oscillating **v** vector radiates a field that interferes destructively with the incident light in the forward direction, indicating that the **v** vector is responsible for absorption. The vertical axis, **w**, represents the population difference between the excited and ground states, i.e., $\mathbf{w} = \rho_{11} - \rho_{00}$. The interaction pathway of the photon-echo signal in the AB sequence is the phase conjugate of $\mathbf{E}_A$ and is followed by $\mathbf{E}_B$, and $\mathbf{E}_B$ (-A+B+B). During the first interaction (−A), the system evolves from the initial state (IS), $\rho_{00}$, into the coherent superposition state between $\rho_{00}$ and $\rho_{11}$, causing the Bloch vector to point along the **u**-axis. During the inter-pulse delay, $\tau$, the system accumulates the phase through coherent evolution until pulse $\mathbf{E}_B$ arrives (final state, FS). The corresponding phase evolution during $\tau$, denoted by $\varphi_\tau$, is represented by the magenta-shaded region in the double-sided Feynman diagram. Following the two successive interactions with $\mathbf{E}_B$ (B + B), the system evolves into the coherent state $\rho_{10}$ while retaining the accumulated phase information, thereby enabling the emission of the photon-echo signal carrying the phase $\varphi_\tau$. Figure 3(b) illustrates the pump-probe (PP) signal, which is characterized by the interaction sequence -A+A+B. During the first two simultaneous interactions with $\mathbf{E}_A$ (-A+A), the system evolves from the initial state (IS), $\rho_{00}$, to the excited-state population, $\rho_{11}$ (FS). Since the Bloch vector in the final state points completely along the north pole, indicating complete population inversion, no phase is accumulated during the inter-pulse delay, leading to $\varphi_\tau = 0°$. It is important to note that the PP signal is irrelevant to the coherent phase evolution, which can be clearly confirmed from $\varphi_\tau = 0°$. After the subsequent interaction with $\mathbf{E}_B$ (B), the system evolves into a state that emits the nonlinear signal carrying the information of the population dynamics.

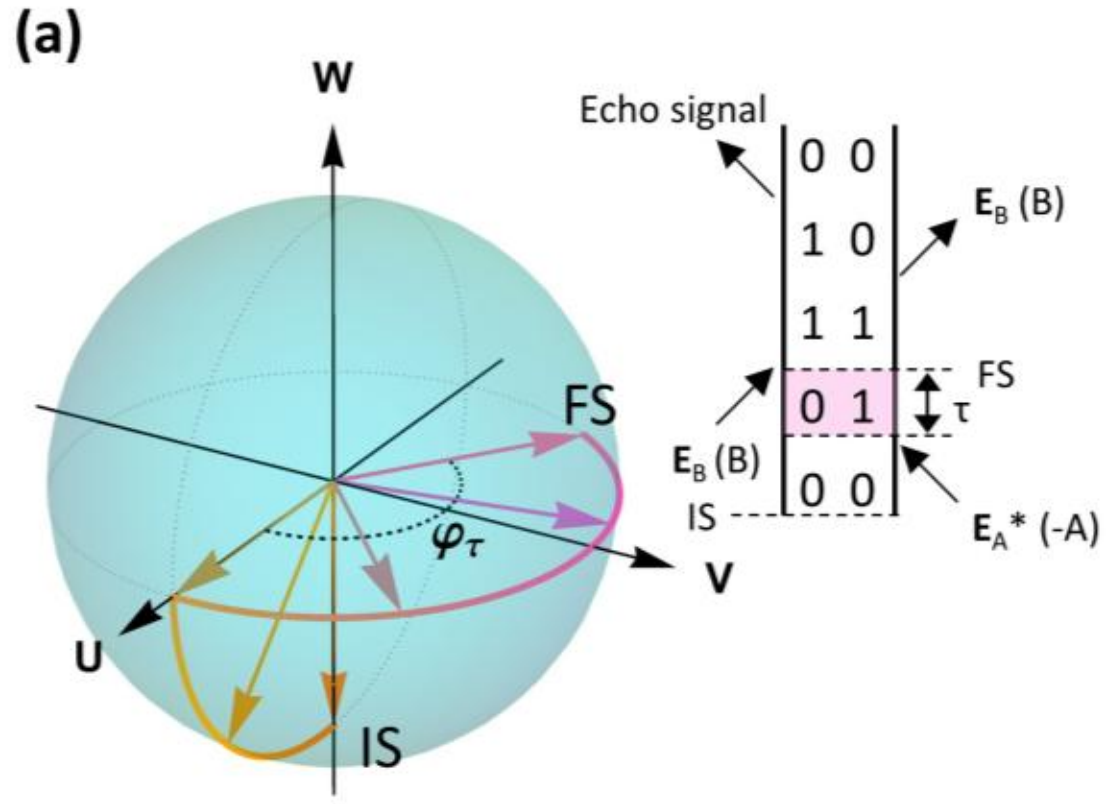

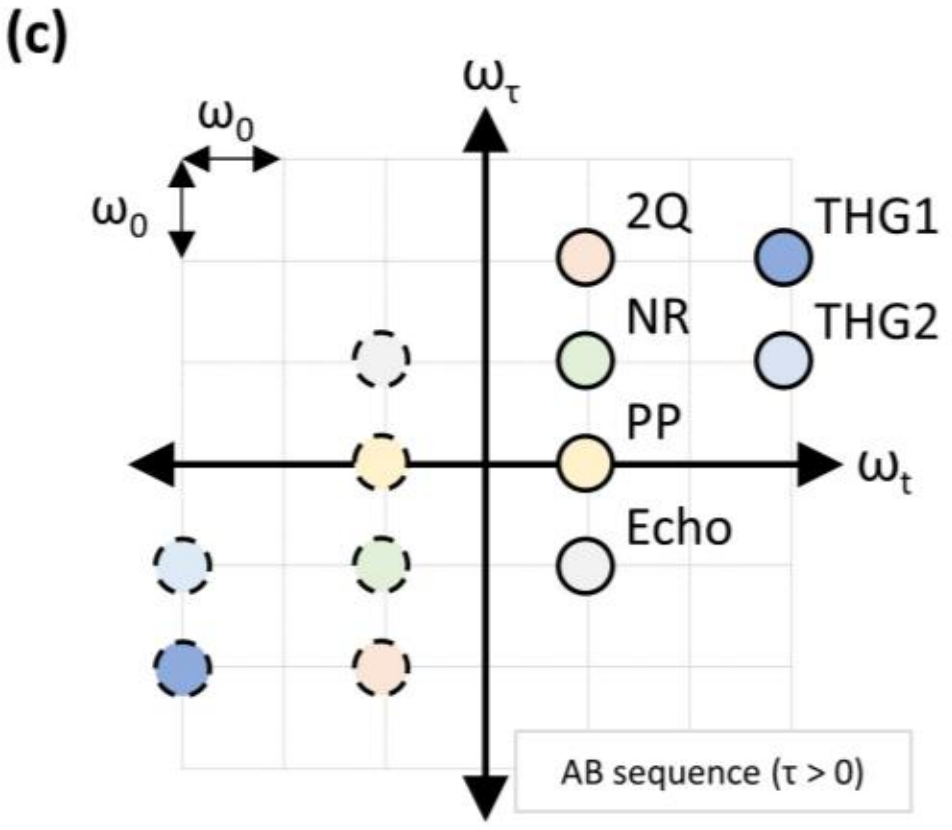

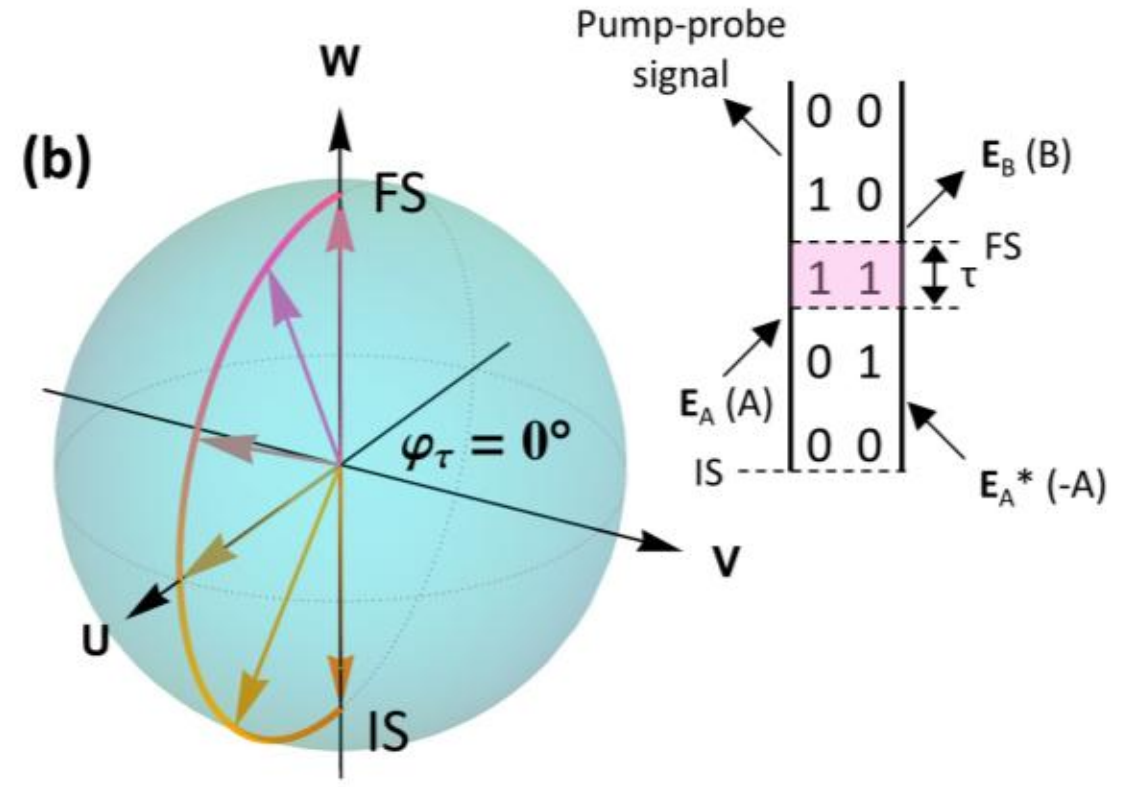

| Label | Signal | Liouville pathway | $(\omega_t, \omega_\tau)$ |
|---|---|---|---|
| Echo | Photon-echo (Rephasing) | -A+B+B | $(\omega_0, -\omega_0)$ |
| PP | Pump-probe | -A+A+B | $(\omega_0, 0)$ |
| NR | Non-rephasing (Free-induction decay) | A-B+B | $(\omega_0, \omega_0)$ |
| 2Q | Second Quantum | A+A-B | $(\omega_0, 2\omega_0)$ |
| THG1 | Third-Harmonic Gen. | A+A+B | $(3\omega_0, 2\omega_0)$ |
| THG2 | Third-Harmonic Gen. | A+B+B | $(3\omega_0, \omega_0)$ |

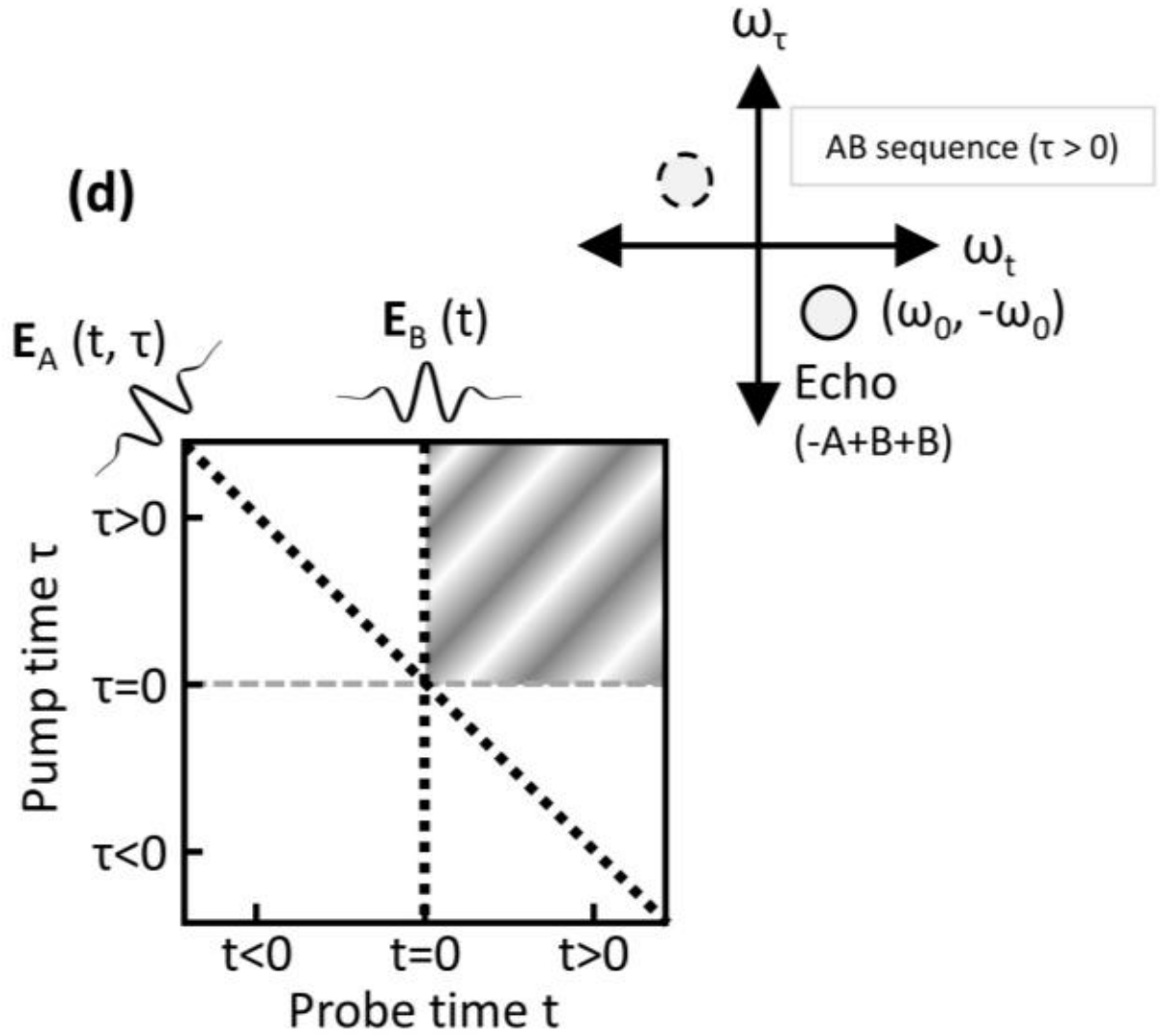

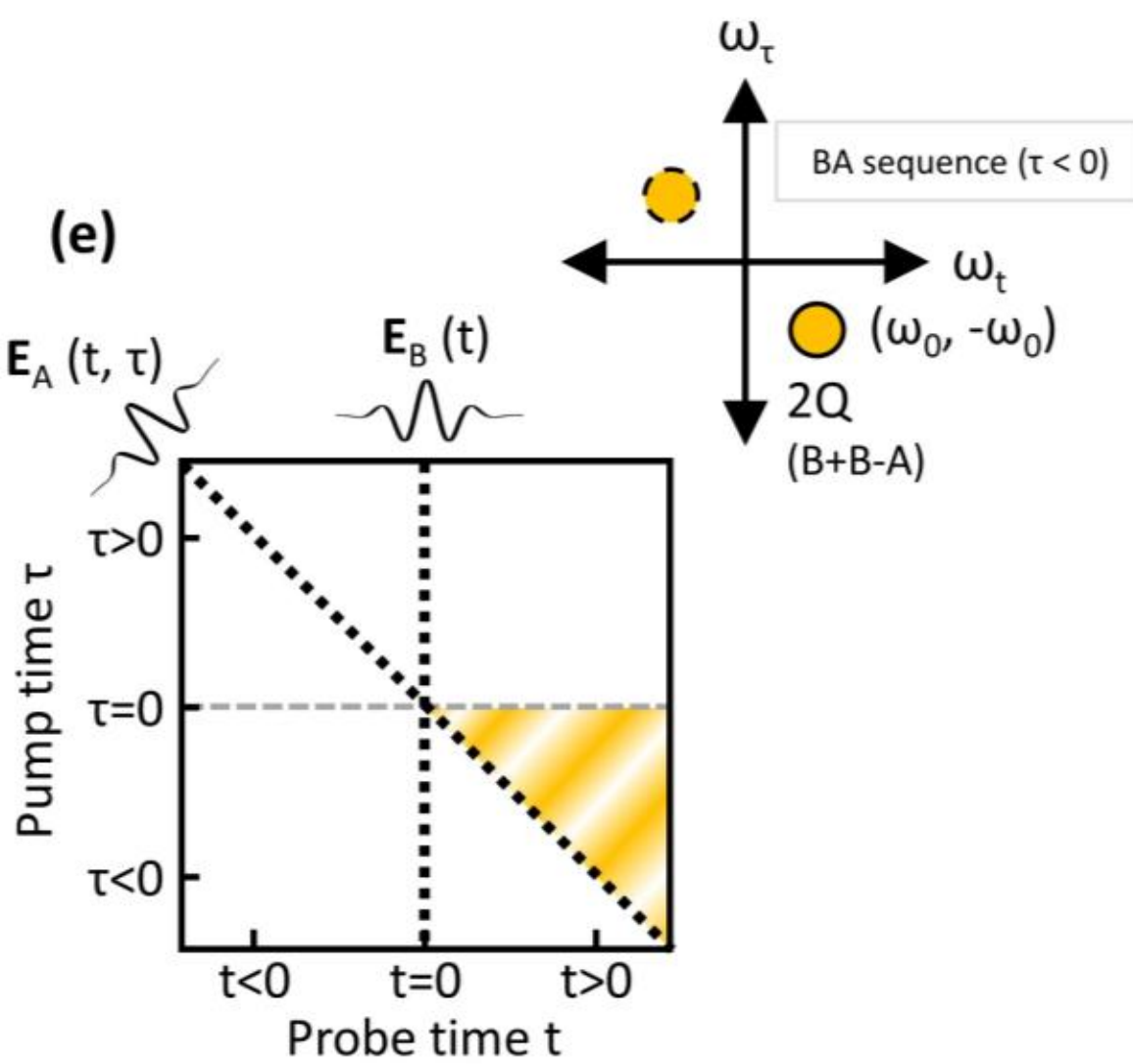

**FIGURE 3 |** Density-matrix representation of quantum evolution and Liouville pathways in THz-2DCS. (a**)** Bloch-sphere representation of the photon-echo (rephasing) signal. The system evolves from the initial state (IS) through the first interaction with $E_A^*$, generating an optical coherence on the equatorial plane of the Bloch sphere. During the time delay $\tau$ between $E_A$ and $E_B$, the coherence accumulates the phase $\varphi_\tau$ until reaching final state (FS)**.** After the subsequent two interactions with $E_B$, the system reaches the state, where the radiating coherence $\rho_{10}$ is generated to emit the photon-echo signal carrying the accumulated phase information. The magenta-shaded region in the Feynman double-sided diagram denotes the coherence-evolution period. (b) Bloch-sphere representation of the pump-probe (PP) signal. The first two interactions with $E_A$ generate the excited-state population $\rho_{11}$, causing the Bloch vector to point toward the north pole (final state). Because the Bloch vector has no transverse (equatorial) component during the delay $\tau$, no phase evolution occurs, resulting in $\varphi_\tau$=0°. (c) The positions of observable third-order nonlinear signals in the frequency domain. Distinct Liouville pathways produce spectrally separated photon-echo (Echo), pump-probe (PP), non-rephasing (NR), second-quantum (2Q), and third-harmonic generation (THG1 and THG2) signals, enabling coherent and population dynamics to be distinguished. The solid circles indicate the frequency positions of the nonlinear signals, while the dotted circles represent their symmetric counterparts arising from the Fourier transform, reflecting the inherent point symmetry of the 2D frequency domain. The table summarizes the corresponding interaction pathways and spectral coordinates. The two-dimensional. The two-dimensional temporal patterns for AB sequency photon-echo signal (d) and BA sequency second quantum signal (e).

The third-order nonlinearities described in Eq. 7 in the frequency domain are summarized in Fig. 3(c). The nonlinearities thereof are determined by the different combinations of the linear combination of $\omega_A$ and $\omega_B$, leading to the different Liouville pathways. For instance, the photon echo signal corresponds to -A+B+B, appearing at ($\omega_0$, -$\omega_0$). Here, $\omega_t$ represents the detection frequency, whereas $\omega_\tau$ is the pump frequency that drives the oscillations of excited quasiparticles in the system. Intuitively, $\omega_t$ corresponds to the observable "color", while $\omega_\tau$ represents the blinking frequency of that color. Therefore, third-harmonic generation (THG) appears at three times the fundamental driving frequency of light (i.e., $\omega_t$=3$\omega_0$). The second-quantum (2Q) signal appears at $\omega_t$= $\omega_0$, enabling it to be observed even in the absence of phase-matching conditions, where this interpretation is consistent with our earlier conclusion. Coherence-related signals, such as photon echo (echo), non-rephasing (NR), and second-quantum (2Q), occur at $\omega_\tau \neq 0$, whereas coherence-independent signals, such as pump-probe (PP), appear at $\omega_\tau$=0. This distinction arises because PP involves resonant excitation, resulting in a significantly longer lifetime compared to parametric (non-resonant) excitations. Assigning these nonlinearities to specific coordinates in the frequency domain allows us to access the underlying individual susceptibilities of low-energy quasiparticles, a capability uniquely offered by THz-2DCS.

The AB sequence refers to the case in which $\mathbf{E}_A$ precedes $\mathbf{E}_B$, so that $\mathbf{E}_A$ ($\mathbf{E}_B$) acts as the pump (probe) beam. At $\tau$=0, the two pulses overlap, and for $\tau$<0, the roles are interchanged, with $\mathbf{E}_B$ becoming the pump, namely, the BA sequence. The 2Q signal in the BA sequence, arising from the initial two interactions with $\mathbf{E}_B$ (B+B) and the following interaction with $\mathbf{E}_A^*$ (-A), is located at $(\omega_t, \omega_\tau) = (\omega_0, -\omega_0)$. The echo in the AB sequence appears at the same frequency position, which makes it challenging to assign the physical origin of the $(\omega_0, -\omega_0)$ peak, whether it stems from the AB echo or the BA 2Q. Inverse Fourier transformation (IFFT), which can transfer the frequency domain back to the time domain, allows the two contributions to be distinguished. By causality, the AB echo must appear for $\tau$>0 and after the probe pulse $E_B$ (t>0), so that its fringe pattern is confined to the first quadrant (Fig. 3(d)). The BA 2Q signal, by contrast, emerges for $\tau$<0 and only after the probe pulse $\mathbf{E}_A$, so that the corresponding pattern occupies the fourth quadrant and is bounded by the causality edge at $t=-\tau$, giving it a triangular shape (Fig. 3(e)). Note that the phase front of the observed fringe pattern is perpendicular to the corresponding 2D frequency vector for each of the nonlinear signals.

### 2.3. Homogeneous vs. Inhomogeneous broadening

For THz-2DCS, the spectral width of the nonlinear signals along a diagonal and anti-diagonal direction provides a microscopic origin of scattering mechanisms. Here, the diagonal direction refers to the direction toward increasing both $\omega_\tau$ and $\omega_t$. The anti-diagonal direction is perpendicular to the diagonal direction. When a quantum material is excited by coherent light, the induced phase coherence among microscopic emitters decays over time due to interactions with the surrounding environment. The total dephasing rate, denoted as $1/T_{total}$, governs the overall decay of quantum coherence and is fundamentally composed of two distinct physical relaxation mechanisms: population decay $T_1$ and population-conserving scattering process $T_2$, as [56,57]:

$$\frac{1}{T_{total}} = \frac{1}{2T_1} + \frac{1}{T_2} \tag{10}$$

$T_1$ originates from inelastic scattering, such as scattering by phonons, whereas $T_2$ results from elastic scattering by impurities. Since the dephasing by $T_1$ is related to the carrier population and resonance excitation, $T_1$ is longer than the coherence-related dephasing $T_2$ in a common case. In other words, the excited carriers relax back to the equilibrium state after the coherence between the ground and excited states has dephased. From the Bloch sphere perspective, the Bloch vector shrinks toward the interior of the Bloch sphere (mixed state) from the surface (pure state) through dephasing. Since $T_1$ ($T_2$) is related (unrelated) to the relative populations of carriers, the trajectory of the Bloch vector is longitudinal (transverse), such that the corresponding decay processes are often called the longitudinal and transverse decays, respectively.

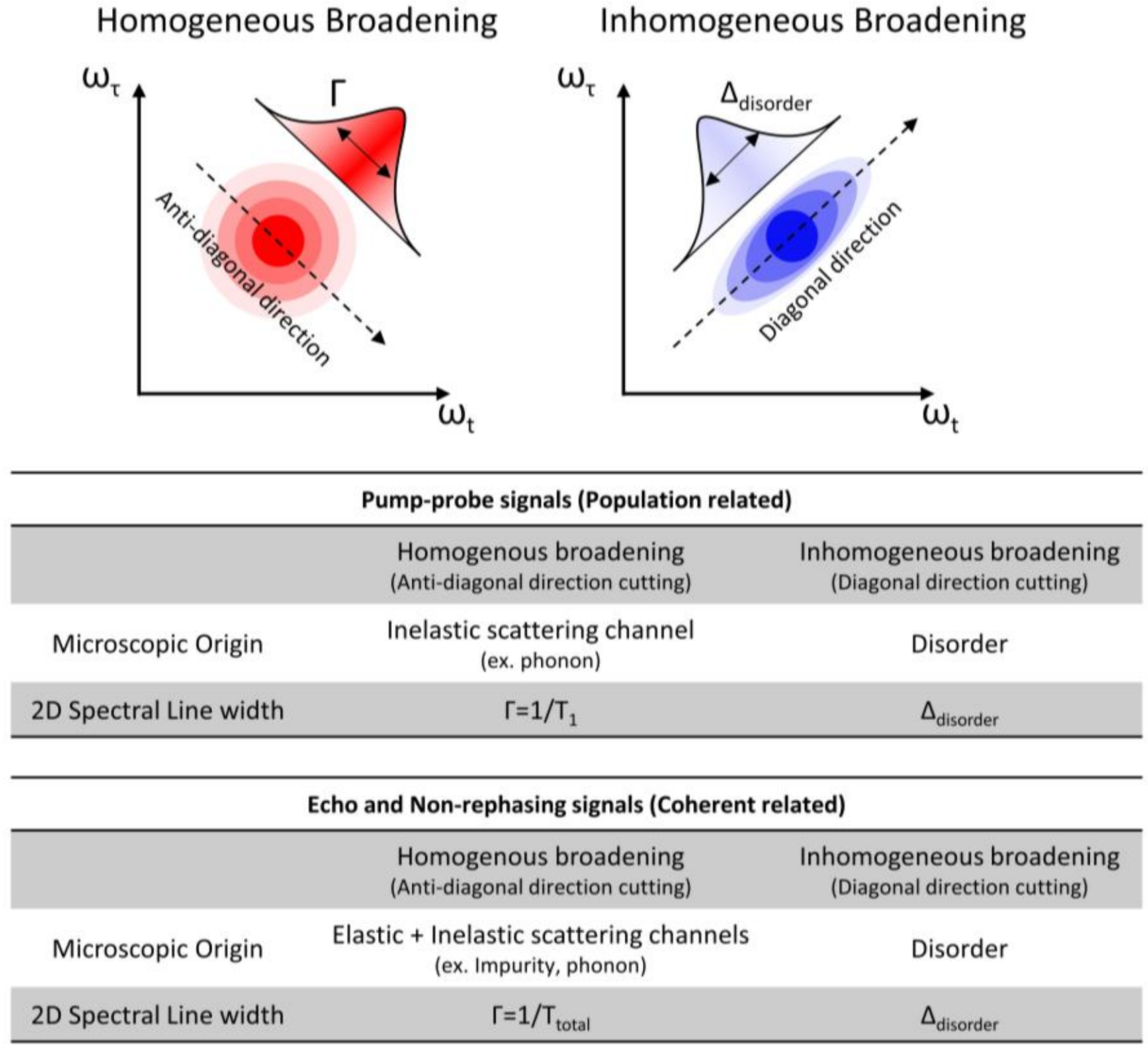

| Pump-probe signals (Population related) | | |
|---|---|---|
| | Homogenous broadening (Anti-diagonal direction cutting) | Inhomogeneous broadening (Diagonal direction cutting) |
| Microscopic Origin | Inelastic scattering channel (ex. phonon) | Disorder |
| 2D Spectral Line width | $\Gamma$=1/$T_1$ | $\Delta_{disorder}$ |

| Echo and Non-rephasing signals (Coherent related) | | |
|---|---|---|
| | Homogenous broadening (Anti-diagonal direction cutting) | Inhomogeneous broadening (Diagonal direction cutting) |
| Microscopic Origin | Elastic + Inelastic scattering channels (ex. Impurity, phonon) | Disorder |
| 2D Spectral Line width | $\Gamma$=1/$T_{total}$ | $\Delta_{disorder}$ |

**FIGURE 4 |** Homogeneous and inhomogeneous broadening**.** Schematic illustration of the spectral linewidths along the anti-diagonal and diagonal directions in two-dimensional spectra. Homogeneous broadening ($\Gamma$) originates from elastic ($1/T_2$) and/or inelastic scattering processes ($1/T_1$) that determine the total coherence dephasing rate ($1/T_{total}$), whereas inhomogeneous broadening ($\Delta_{disorder}$) arises from static disorder in the sample. The linewidths of population- and coherence- related nonlinear signals provide different dephasing information.

Figure 4 shows the comparison of homogeneous and inhomogeneous broadening, observable in the frequency domain of THz-2DCS. Homogeneous broadening originates from the coupling of THz-excited quasiparticles ($\omega_\tau$) to other degrees of freedom, such as phonons and other scattering channels. Through these interactions, the energy of excited quasiparticles is dissipated into the surrounding environment (=strong coupling), resulting in the inconsistency energy between the pump frequency ($\omega_\tau$) and the probe frequency ($\omega_t$), which manifests as homogeneous linewidth broadening. The pump-probe (PP) signal reflects the population dynamics, whereas the photon echo (Echo) and non-rephasing (NR) signals originate from phase-sensitive nonlinear processes. Thus, the widths Γ of PP and phase-related signals are related to $1/T_1$ and $1/T_{total}$, respectively, suggesting that one can extract pure dephasing time $T_2$ by subtracting the widths of phase-related signals from PP. The linear THz spectroscopy, e.g., THz-TDS, only allows us to approach the total relaxation rate $1/T_{total}$ by utilizing proper modeling such as the Drude formalism. In this aspect, the coherent-related carrier dynamics can be uniquely investigated by THz-2DCS.

In contrast to homogeneous processes, inhomogeneous broadening arises from static spatial structural disorder of the sample. Because of the disorder resulting in a variation of a local crystal environment, the excitation frequency for the quasiparticles is slightly detuned, and hence the width in the frequency domain is elongated along the diagonal direction, i.e., $\omega_t=\omega_\tau$. The degree of static disorder is determined by the structural quality of the sample, including defects, impurities, and lattice imperfections. Because disorder universally gives rise to the elongation of the nonlinear signal, the signal width along the diagonal direction (=inhomogeneous broadening, $\Delta_{disorder}$) reflects the strength of disorder in the sample. The microscopic origins and physical interpretations of the spectral linewidths associated with different nonlinear signals, including population- and coherence-related signals, are summarized in the table shown below in Fig. 4.

### 2.4. Rabi oscillation (non-perturbative region)

Under moderate THz electric fields (~kV/cm), the light–matter interaction can be well described by a perturbative expansion of the polarization in powers of the electric field (Eq. (1)), where higher-order $\chi^{(n)}$ terms provide progressively smaller corrections. In previous sections, we introduced nonlinear signals, mostly in third-order nonlinear processes, described by this perturbative expansion. As the THz electric field approaches the MV/cm scale, however, the light-matter interaction can no longer be regarded as a small correction to the unperturbed system. To describe this non-perturbative regime properly, the full time-dependent Hamiltonian must be treated explicitly rather than using a perturbative expansion. The simplest and most representative model for such non-perturbative phenomena is a resonantly driven two-level system consisting of a ground state and an excited state. We can obtain the intuitive understanding of the population and coherent dynamics in this simplest system by solving the full time-dependent Hamiltonian, and the periodic population exchange is known as Rabi oscillation [58-60]. Thus, one can understand the non-perturbative nonlinear process observed in THz-2DCS by implementing a Rabi oscillation via interaction with two intense THz electric fields.

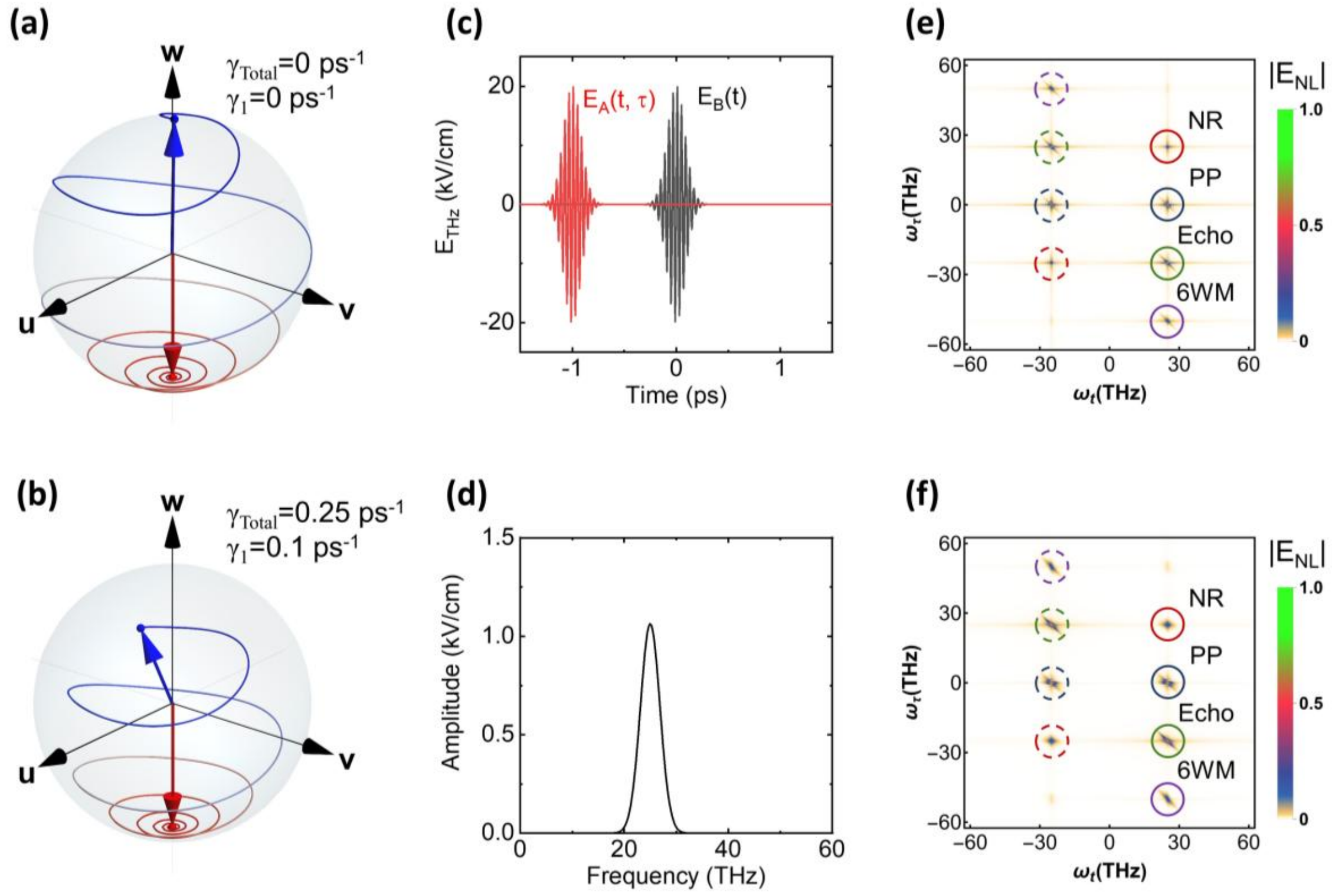

**FIGURE 5 |** The population dynamics of carriers in a two-level system. The Bloch vector trajectories of the two-level system without damping ($\gamma_{total} = \gamma_1 = 0$ s$^{-1}$) (a) and with damping ($\gamma_{total} = 0.25$ ps$^{-1}$ and $\gamma_1 = 0.1$ ps$^{-1}$) (b). The initial state is set to the ground state, pointing to the south pole ($\mathbf{w} = -1$). Time evolution is indicated by a color gradient from red to blue. (c) Incident THz electric field recorded in the time domain. $\mathbf{E}_A$ and $\mathbf{E}_B$ are identical. (d) Corresponding amplitude spectrum of $\mathbf{E}_A$ and $\mathbf{E}_B$. The simulated two-dimensional spectra of Rabi oscillations in a two-level system for without damping ($\gamma_{total} = \gamma_1 = 0$ s$^{-1}$) (**E**) and with damping ($\gamma_{total} = 0.25$ ps$^{-1}$ and $\gamma_1 = 0.1$ ps$^{-1}$) (f).

The underlying origin of Rabi oscillation lies in absorption and stimulated emission, which produce light with properties exactly identical to those of the incident light, including its polarization and propagation direction. By using the analogy with as a classical damping model, one can phenomenologically derive the governs equations for Rabi oscillation, which is known as the optical Bloch equations, as [56,61-64]:

$$\partial_t \mathbf{u} = -\omega_{21}\mathbf{v} + 2\Omega_i \mathbf{w} - \gamma_{total}\mathbf{u} \quad (11)$$

$$\partial_t \mathbf{v} = \omega_{21}\mathbf{u} + 2\Omega_r \mathbf{w} - \gamma_{total}\mathbf{v} \quad (12)$$

$$\partial_t \mathbf{w} = -2\Omega_i \mathbf{u} - 2\Omega_r \mathbf{v} - \gamma_1 \mathbf{w} + \Gamma_{12} \quad (13)$$

where **u**, **v**, and **w** are the Bloch vectors, and $\Omega = \Omega_r + i\Omega_i$ is the complex Rabi frequency, which is given by is given by $\mu_{eg} \cdot E_0/\hbar$. Here, $\mu_{eg}$ is the transition electric-dipole moment and $E_0$ is the amplitude of the applied electric field. $\gamma_{total}$ denotes the total dephasing rate, defined by $\gamma_{total}=1/T_{total}$, and $\gamma_1 = 1/T_1$ accounts for the dephasing related to the population-decay scattering mechanism (see. Eq. (10)). The term $\Gamma_{12}$ represents an incoherent net population from the ground state and the excitation state. According to Einstein coefficients, an excited carrier can return spontaneously to a lower-energy equilibrium state without external optical stimulation driven by the incident light. As a result, two emission processes, spontaneous and stimulated emission, contribute to the decay dynamics, which are incoherent and coherent with respect to the incident light. To describe spontaneous emission, $\gamma_1$ and $\gamma_{total}$ are phenomenologically inserted in the optical Bloch equations (Eqs. (11)-(13)).

Figure 5 shows the population dynamics of carriers in a two-level system, as obtained from the optical Bloch equations. The Rabi frequency $\Omega$ is approximately 7 THz, based on a maximum THz electric-field amplitude of $E_0$ = 20 kV/cm and a transition dipole moment of $\mu_{eg} = 3.7 \times 10^{-28}$ C·m. Note that the 20 kV/cm electric field strength is achievable by employing non-organic crystals, such as GaP or ZnTe, combined with the low-repetition-rate amplifier femto-second laser system. The electric-field profiles for $\mathbf{E}_A$ and $\mathbf{E}_B$ are shown in Fig. 5(c), and the amplitude profile is presented in Fig. 5(d). $\mathbf{E}_A$ and $\mathbf{E}_B$ are set to identical pulses with a center frequency of $\omega_0$ = 25 THz, which can practically be generated by GaSe crystal. When $\gamma_{total} = 0$ $s^{-1}$ and $\gamma_1 = 0$ $s^{-1}$, the Bloch vector evolves from $\mathbf{w} = -1$ (the south pole) to $\mathbf{w} = 1$ (the north pole). Its time evolution is indicated by the color gradient from red to blue. Throughout this evolution, the Bloch vector remains on the surface of the Bloch sphere, indicating that coherence is conserved and the state remains pure: $\mathbf{u}^2 + \mathbf{v}^2 + \mathbf{w}^2 = 1$ (see Fig. 5(a)). Upon turning on the damping as $\gamma_{total} = 0.25$ $ps^{-1}$ and $\gamma_1 = 0.1$ $ps^{-1}$, spontaneous-emission-induced emission comes into play for the carrier population dynamics. In this case, the Bloch vector does not reach complete population inversion ($\mathbf{w}$ = 1) and loses coherence, so that it lies inside the Bloch sphere, corresponding to a mixed state: $\mathbf{u}^2 + \mathbf{v}^2 + \mathbf{w}^2 < 1$ (see Fig. 5(b)). It is important to note that $2\gamma_{total}$ should be smaller than $\gamma_1$ (see Eq. (10)).

Figures 5(e) and 5(f) show simulated two-dimensional spectra of Rabi oscillations in a two-level energy system under conditions corresponding to those in Figs. 5(a) and 5(b), respectively. These spectra were obtained by performing a two-dimensional Fourier transform using only the $\tau$>0 data, corresponding to the AB pulse sequence; namely, $\mathbf{E}_{NL}|$ was set to zero for the case of $\tau$<0. Owing to the long wavelength of the THz radiation, the dipole distribution within the THz focal region can be approximated as a planar dipole distribution rather than a point dipole. Under the condition that the THz wavelength is much larger than the sample thickness, the emitted THz electric field $\mathbf{E}_{THz}(t)$, induced by the incident THz field $\mathbf{E}_0(t)$, is calculated by [65]:

$$E_{THz}(t)=E_0(t)-\alpha\partial_t\mathbf{u} \tag{14}$$

$\alpha$ is defined by $\mu_0 c/2$. $\mu_0$ is the vacuum magnetic permeability, and $c$ is the speed of light. In our simulation, we employed $\alpha=5.56\cdot10^{-15}$ $s^{-1}$.

The several noticeable features in our simulation can be observed, giving a deep insight for the nonlinear carrier dynamics. We consider the two-level system with a nonvanishing transition dipole moment operator $\mu$ from the ground state $|g>$ to the excited state $|e>$, $\mu_{eg} = <e|\mu|g>$, while the diagonal matrix elements $\mu_{gg} = <g|\mu|g>$ and $\mu_{ee} = <e|\mu|e>$ are assumed to vanish. These diagonal elements correspond to the permanent dipole moments of the ground and excited states, respectively, rather than transition dipoles. Under this assumption, the induced polarization contains only odd powers of the driving field, giving rise exclusively to odd-order nonlinear signals, such as third-order nonlinear signals (four-wave mixing, 4WM) and fifth-order nonlinear signals (six-wave mixing, 6WM). Four peaks are clearly discernible, and we assigned the origin of the nonlinear process for each peak. Three of the processes (NR, PP, and Echo) originate from third-order nonlinearity, while the remaining one arises from fifth-order nonlinearity: -A-A+B+B+B. Given that the photon echo signal for the AB sequence is -A+B+B, this fifth-order nonlinear signal can be understood as the result of adding -A+B to the AB photon echo signal on the front and back sides, giving rise to the six-wave mixing (6WM) signal appearing at ($\omega_0$, $-2\omega_0$). Importantly, one can clearly confirm that the 6WM signal in the spectrum provided in Figs. 5(e) and (f), meaning that the Rabi oscillation represents the non-perturbative nonlinear responses.

As the scattering rate $\gamma$ increases, equivalent to a decrease in the total coherence time T, the linewidth generally becomes broader, since faster dephasing suppresses the resonant excitation efficiency and shortens the effective interaction time contributing to the signal. As discussed in the previous section 2.3, the line-cut analysis along the diagonal and anti-diagonal directions for the coherence-related signals, e.g., Echo, manifests a different broadening profile compared to the coherence-unrelated signals, i.e., PP. When $\gamma_1$ increases, which is related to population decay, the peak width is expected to exhibit homogeneous broadening regardless of the type of nonlinear signal, whether it is related to coherence or not. This is the underlying origin showing the observed homogeneous broadening for all nonlinear signals in Figs. 5(e) and (f). Interestingly, the Echo signal exhibits more pronounced homogeneous broadening than the PP signal. This can be understood by the same underlying mechanism; because the Echo signal is the coherent-related nonlinear process, its total dephasing rate $\gamma_{total}$ is increased by the additional contribution from the population-preserved scattering, i.e., $\gamma_2$, on top of $\gamma_1$. Since the PP signal lacks this coherence contribution, its broadening reflects $\gamma_1$ alone, making the width of the Echo signal along the anti-diagonal direction increase more sensitively. Although a detailed analysis is beyond the scope of the present work, the line-cut analyses of the individual spectral features observed in the Rabi oscillation are worth addressing in a future study to get a comprehensive understanding of nonlinear carrier-population dynamics accessible through THz-2DCS.

## 3. THz-2DCS on quantum materials for the investigation of nonlinear dynamics

In previous sections, we introduced knowledge about nonlinear and quantum optics, which are the basic pillars for understanding THz-2DCS. We now move on to the relevant studies in which THz-2DCS is employed as the core technique to investigate the coherent nonlinear dynamics of low-energy excitations emerging in a wide range of quantum materials.

### 3.1. Nonlinear Plasmon dynamics

Plasma is a neutral gas composed of charged particles [66]. In solid-state materials, free carriers can be regarded as an electron plasma embedded in a positively charged background, and their collective longitudinal oscillations are quantized as plasmons. Because the low energy of THz radiation predominantly drives intraband transitions rather than interband excitations, the plasma in solid-state systems can be coherently and efficiently accelerated by THz fields, allowing THz light to serve as an effective probe light for investigating the nature of plasmons. The plasmonic motion generally persists in an environment where a terahertz field (THz) is present; therefore, its characteristic response time can be tailored by controlling the duration of the THz pulse [2]. This feature distinguishes plasmonic nonlinearity from other nonlinear mechanisms, such as thermal nonlinearity [2], which can persist even after the THz field has been terminated. Therefore, nonlinear plasmon dynamics offer a promising platform for ultrafast optical switching and modulation [7].

Nonlinear plasmon dynamics were investigated in narrow-gap InSb, exhibiting the plasma edge $\nu_p$ near 2.2 THz and the LO-phonon mode near 5.7 THz (Fig. 6(a)) [67]. The two-dimensional spectra displayed pronounced features at ($\omega_t$=$\nu_p$, $\omega_\tau$=0) and ($\nu_p$, 2$\nu_p$), as well as phonon-related features at ($\nu_{LO}$, 0) and ($\nu_{LO}$, 2$\nu_p$). As discussed in Section 2.2, $\omega_t$ and $\omega_\tau$ represent the detection and excitation frequencies, respectively. The feature at $\omega_t$=$\nu_p$ indicates that the emitted nonlinear THz signal oscillates at the plasma frequency. Its dependence on the interpulse delay $\tau$ is encoded in $\omega_\tau$. Therefore, the origin of a nonlinear signal can be assigned primarily from its detection frequency $\omega_t$. In this case, the peaks at ($\nu_p$, 0) and ($\nu_p$, 2$\nu_p$) are attributed to nonlinear plasmon dynamics, as both emit at the plasma frequency, with their amplitude oscillating at 0 and 2$\nu_p$ frequencies, respectively. The frequency window of the applied THz light is within 15 THz, which is far below the InSb band gap energy (~ 50 THz), such that the nonlinear mechanism related to interband excitation is excluded. One possible origin of the nonlinear signals appearing at $\omega_t$=$\nu_p$ is ballistic intraband acceleration of carriers into the nonparabolic conduction band, resulting in the modification of the effective carrier mass, and therewith the nonlinear modification of $\nu_p$. A second possible contribution is a field-induced change of carrier density, for example, through impact ionization under intense THz driving. The modulation of the plasma frequency also modifies the coupling between the plasmon and LO phonon, producing phonon-related nonlinear signals. To distinguish the two mechanisms of nonlinear plasmonic response, parity-resolved THz-2DCS was used by reversing the polarity of the nonlinear THz field emitted and separating the nonlinear response into even- and odd-parity components. This analysis helps distinguish predominantly incoherent carrier generation, which changes the carrier density, from phase-coherent ballistic intraband motion. The odd-parity peak at ($\nu_p$, $\nu_p$) was attributed to ballistic electron acceleration in the nonparabolic conduction band, where the local band curvature and thus the effective mass are dynamically modified.

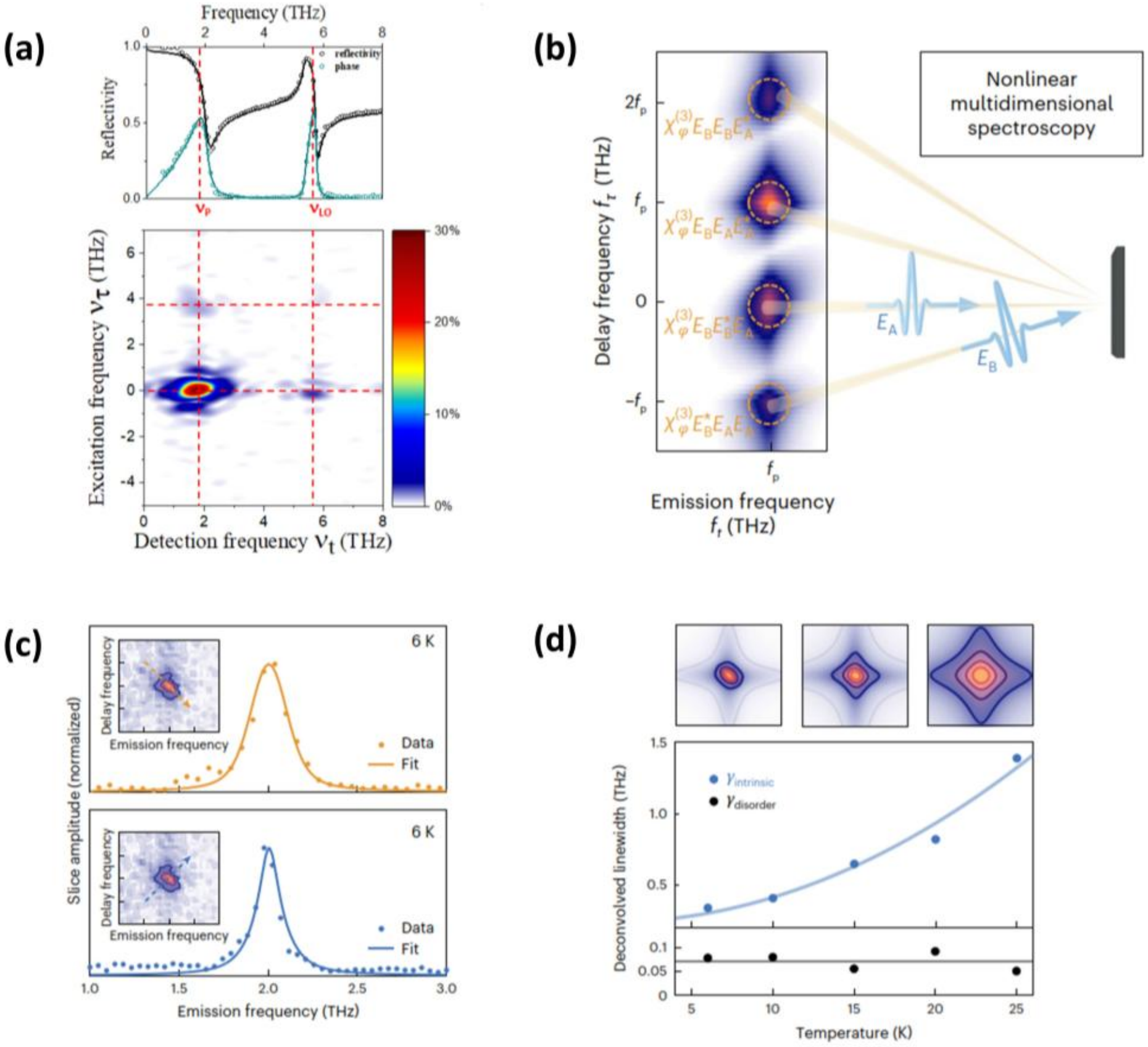

**FIGURE 6 |** (a) Reflectivity amplitude (black) and phase (green) of InSb (upper panel) and the experimentally obtained nonlinear 2D spectra (lower panel). Adapted from Ref. [67] with permission from Optica Publishing Group, copyright (2019). (b) Schematic diagram for non-collinear THz-2DCS. (c) Line-cut analysis of the Josephson-echo signal along the diagonal (upper panel, orange) and anti-diagonal (lower panel, blue) directions. (d) Temperature dependence of the scattering rate γ, separated into intrinsic and electronic-disorder contributions. The intrinsic contribution is associated with the anti-diagonal linewidth, whereas the disorder contribution is extracted from the diagonal line cut. Graphs in (b), (c), and (d) are reprinted from Ref. [68], with permission from Springer Nature, copyright (2024).

Josephson plasmons, i.e., collective oscillations of superconducting carriers between adjacent superconducting layers, have also been investigated using THz-2DCS in doped $La_{1.83}Sr_{0.17}CuO_4$ [68]. Because bulk layered superconductors can be opaque at THz frequencies, conventional transmission-based THz-2DCS is often impractical for probing their nonlinear responses. To overcome this limitation, reflection-based THz-2DCS in a non-collinear geometry was employed, enabling the spatial separation and isolation of phase-matched nonlinear signals (Fig. 6(b)). Figure 6(c) presents line-cut analyses along the diagonal (orange) and anti-diagonal (blue) directions of the 2D spectrum (see Section 2.3). Note that the homogeneous broadening along the anti-diagonal direction arises from both inelastic and elastic scattering processes, whereas the inhomogeneous broadening along the diagonal direction is attributed primarily to the static disorder. As the temperature increases, the homogeneous linewidth becomes approximately three times larger than the inhomogeneous linewidth (Fig. 6(d)). This result indicates that electronic static disorder makes only a minor contribution to the interlayer tunneling response between adjacent superconducting layers. This study demonstrates that THz-2DCS, through line-cut analysis of the two-dimensional spectral line shape, can disentangle homogeneous and inhomogeneous contributions to the Josephson-plasmon response. It therefore provides access to the scattering mechanisms of superconducting carriers beyond what is available from conventional linear THz spectroscopy.

### 3.2. Nonlinear Magnon dynamics

The quantized collective motion of spin degrees of freedom is called a magnon. Compared with charge-based excitations, spin flipping requires relatively low energy and can occur on ultrafast femtosecond and/or picosecond timescales. These properties make magnons promising information carriers for next-generation storage and processing devices. In antiferromagnets and ferrimagnets, magnon dynamics, including relaxation and switching, typically occur on picosecond timescales, corresponding to energies of a few meV [69]. Since 1 THz radiation corresponds to approximately 4 meV, THz light is particularly well suited for investigating ultrafast magnon dynamics. At the same time, the low-energy nature of magnons enables them to couple strongly to other excitations in solid-state materials with comparable energy scales, such as phonons. For example, $E_u$ mode phonon lies in 1.5 THz in $MnBi_2Te_4$ [70]. However, investigating the detailed mechanisms of such coupling using conventional one-dimensional spectroscopy is inherently challenging; Spectral features of the two excitations often overlap along the frequency axis, making them difficult to distinguish. In this context, THz-2DCS can be used to characterize magnon coupling dynamics [71]. By expanding the spectroscopic dimension from the probe frequency $\omega_t$, to the pump frequency $\omega_\tau$, THz-2DCS can separate spectral information even when the coupled excitations lie in similar energy ranges [72].

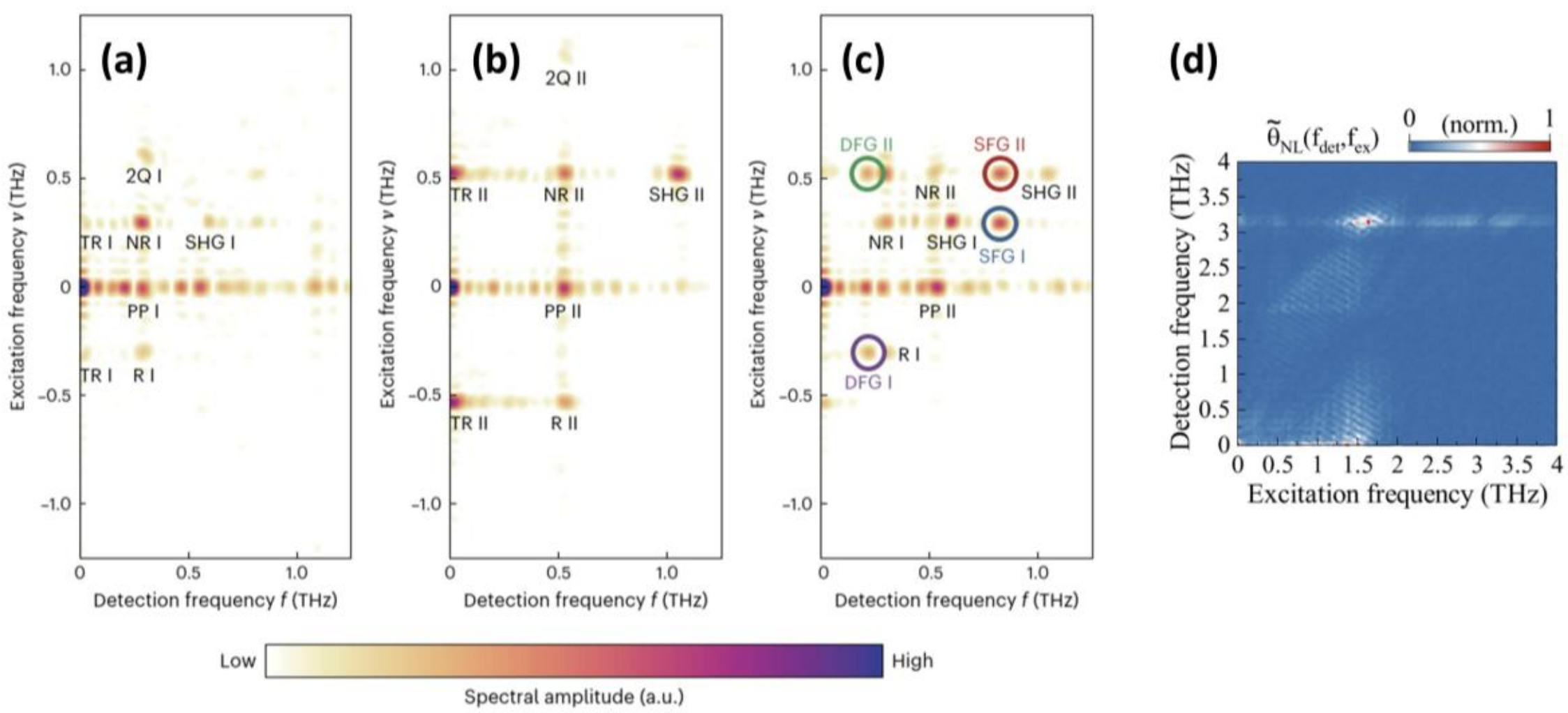

**FIGURE 7 |** 2D spectrum of qFM (a) and qAFM (b), and qFM+qAFM (c) modes of $YFeO_3$**.** The abbreviations in (a)-(c) are provided in the main text. Adapted from Ref. [73] with Springer Nature, copyright (2024). (d) 2D spectrum in $MnBi_2Te_4$ as a function of the excitation frequency ($\omega_\tau$) and the detection frequency ($\omega_t$). The graph in (d) is reprinted from Ref. [70], with permission from American Physical Society, copyright (2023).

Using THz-2DCS, the coherent coupling mechanism between the two magnon modes in $YFeO_3$, i.e., the quasi-antiferromagnetic (qAFM, 0.53 THz) and quasi-ferromagnetic (qFM, 0.30 THz) modes, was unveiled in 2024 [73]. Figures 7(a) and 7(b) show the nonlinear 2D spectra of the qFM and qAFM modes, respectively. Considering the wavevector for the incident THz pulses $\mathbf{E}_A$ and $\mathbf{E}_B$, $\omega_A = (\omega_t = \omega_q, \omega_\tau = \omega_q)$ and $\omega_B = (\omega_q, 0)$, all peaks in Figs. 7(a) and 7(b) can be assigned with reference to Fig. 3(c). Here, the subscript q denotes either the qFM or qAFM mode; for example, $\omega_{qFM} = 0.30$ THz. Each mode can be selectively excited by adjusting the relative angle between the $YFeO_3$ crystal and the polarization of the incident THz electric field. Interestingly, mixed nonlinear signals involving both qFM and qAFM modes appear as DFG I, DFG II, SFG I, and SFG II (see Fig. 7(c)). Here, DFG and SFG denote difference-frequency generation and sum-frequency generation, respectively, while I and II indicate the two possible assignments of the qFM and qAFM modes to $\omega_A$ and $\omega_B$. For case I, $\omega_A$ originates from the qFM mode and $\omega_B$ from the qAFM mode, while for case II, the assignments are reversed. SFG I and SFG II are assigned to $\omega_A + \omega_B$ and $\omega_B + \omega_A$, respectively. By taking $\omega_A = (0.3 \text{ THz}, 0.3 \text{ THz})$ and $\omega_B = (0.5 \text{ THz}, 0 \text{ THz})$,

SFG I and DFGI are expected to appear at (0.8 THz, 0.3 THz) and (0.2 THz, -0.3 THz) in the 2D spectrum. In a similar manner, SHG II and DFG II correspond to peaks at (0.8 THz, 0.5 THz) and (0.2 THz, 0.5 THz), respectively. As can be seen in Fig. 7(c), the peak locations obtained from the experiment are in good agreement with the expectation from combinations of the qFM and qAFM frequencies. This fact demonstrates that they originate from second-order coherent coupling between the qFM and qAFM magnon modes, showing THz-2DCS can be used as for the investigation of the coupling mechanism.

### 3.3. Nonlinear Phonon dynamics

Coupling between the quasiparticles provides a pathway for exciting a Raman-active phonon that cannot be directly driven by the infrared (IR) beam. In $MnBi_2Te_4$, the Raman-active $E_g$ phonon is located at 3.14 THz, and the IR-active $E_u$ phonon is at 1.47 THz [70]. Figure 7(d) shows the two-dimensional spectrum obtained by THz-2DCS on In $MnBi_2Te_4$. Two prominent peaks appear at the same detection frequency, $\omega_t$=3.14 THz, but at distinct excitation frequencies, $\omega_\tau$=1.47 and 1.67 THz, leading to spectral peaks at (3.14 THz, 1.47 THz) and (3.14 THz, 1.67 THz). The fact that $\omega_t$ coincides with the $E_g$ phonon frequency indicates that the Raman-active phonon is excited through a nonlinear excitation pathway. In particular, the peak at (3.14 THz, 1.47 THz) suggests that the $E_u$ phonon serves as an excitation source for the $E_g$ phonon, since $\omega_\tau$=1.47 THz exactly matches the $E_u$ phonon frequency. Furthermore, the frequency difference between the $E_g$ and $E_u$ modes is 3.14−1.47=1.67 THz, accounting for the second peak at (3.14 THz, 1.67 THz). These observations support a two-step excitation process; the intense incident THz field first excites the IR-active $E_u$ phonon, and the $E_u$ phonon subsequently couples to and excites the Raman-active $E_g$ phonon. This excitation pathway is referred to as the photo-phononic mechanism. Consequently, the coupling mechanism between quasi-particles can be scrutinized by using THz-2DCS.

### 3.4. Nonlinear carrier dynamics in non-perturbative region

As discussed in Section 2.4, nonlinear signals cannot be explained based on an expansion of the electric susceptibility; hence, higher-order nonlinearities can be observed simultaneously with intensities comparable to those of lower-order nonlinearities. Quantum wells are suitable material platforms for investigating these non-perturbative nonlinearities using THz-2DCS, since their energy levels can be engineered by adjusting their physical dimensions to energetically match the incident THz light. Namely, the interband excitation strength, determined by the transition dipole moment between quantum-well energy levels [65,74-77], is very large, allowing access to this non-perturbative interaction regime with moderate THz electric fields of approximately 10 kV/cm. One of the representative works for Nonlinear carrier dynamics in the non-perturbative region is the strongly coupled intersubband polaritons in semiconductor quantum wells [76]. The system is characterized by a vacuum Rabi frequency, which determines the polaritonic mode splitting. Under strong-field THz excitation (~55 kV/cm), the response enters a non-perturbative regime, where the four- (4WM) and six-wave mixing (6WM) signals emerge together (Fig. 8(a)), which can be well reproduced by the simulation based on the Rabi oscillation (Fig. 8(b)). This behavior, which a simple perturbative susceptibility expansion cannot describe, arises from strong-field-driven saturation of the intersubband transition.

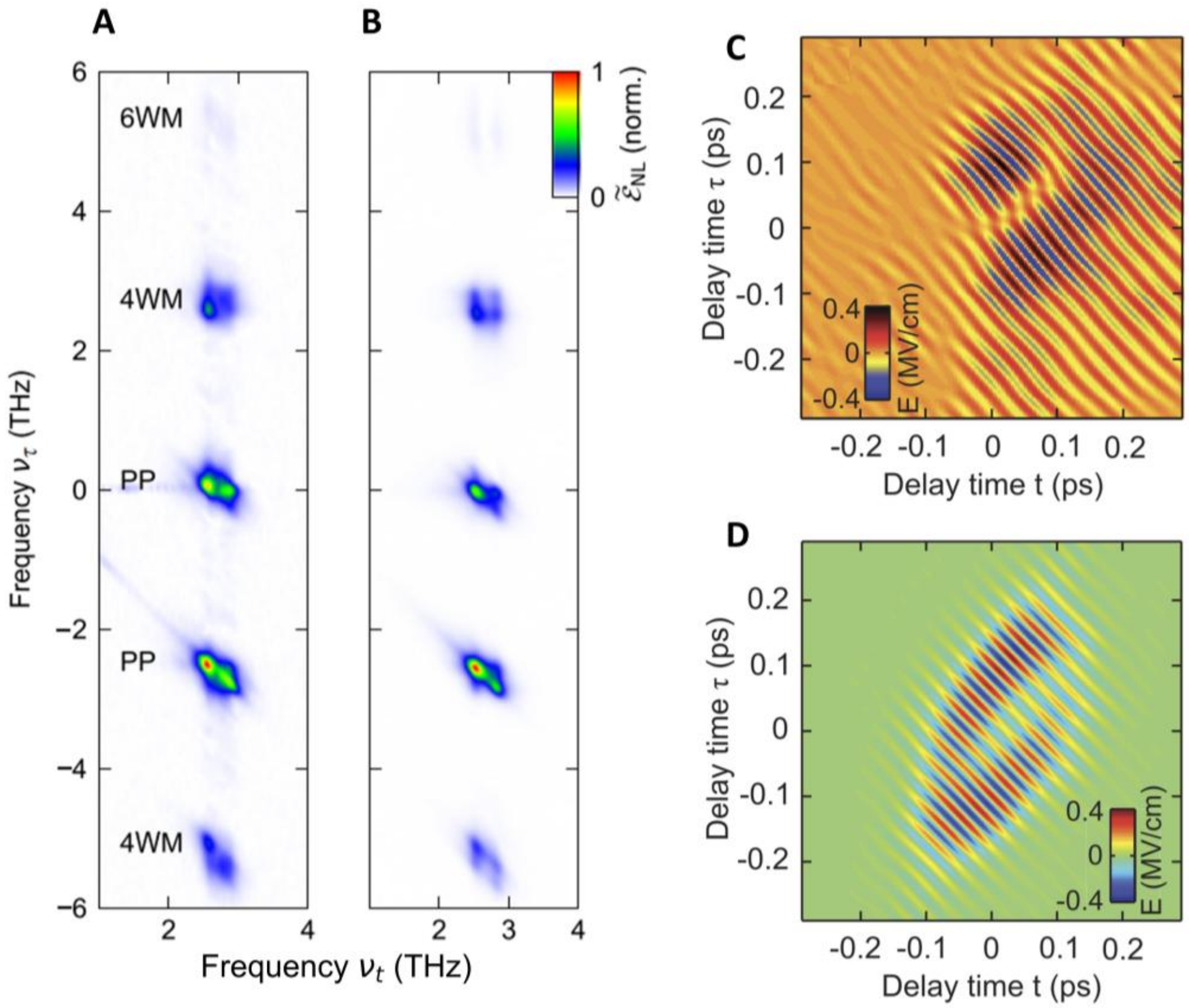

**Fig. 8.** Experimentally obtained 2D spectrum of the multi-quantum well structure consisting of 35 GaAs quantum wells (**A**) and the corresponding simulation spectrum (**B**). Adapted from Ref. [76] with Springer Nature, copyright (2020). (**C**) Experimentally obtained 2D spectrum from bulk (100) InSb with the electric field strength of the THz light about 5 MV/cm. (**D**) Corresponding 2D spectrum simulated with the optical Bloch equations. Graphs in (**C**) and (**D**) are reprinted from Ref. [78], with permission from American Physical Society, copyright (2012).

Although Rabi oscillations are conventionally formulated for an ideal two-level system, they provide a useful framework for understanding the interaction between intense THz fields and semiconductors [78]; a 30-μm-thick, undoped (100) InSb single crystal at room temperature was excited by phase-stable, few-cycle THz transients generated through difference-frequency mixing of near-infrared pulse trains in a 370-μm-thick GaSe crystal. The THz pulses had a center frequency of 23 THz, a bandwidth of 8 THz, and peak electric fields ranging from 2 to 5.3 MV/cm. Since the interband transition associated with the InSb band gap lies at around 41 THz, the excitation was strongly off-resonant with a detuning energy of approximately 18 THz. Figures 8(c) and (d) compare the experimental (Fig. 8(c)) and simulated (Fig. 8(d)) two-dimensional time-domain four-wave mixing signals (FWM) obtained by THz-2DCS spectroscopy with 5 MV/cm as the strength of the incident THz electric field. The simulation, based on the optical Bloch equations (see section 2.4) for a driven two-level system, reproduces the splitting of the FWM signal, resulting in the 'S-shape'. For transitions away from the band edge, the interband transition energy increases further, leading to a larger detuning from the driving field. As a result, the nonlinear response is predominantly governed by near-band-edge transitions with the smallest detuning, allowing the continuum response to be well approximated by an effective two-level system. This fact is the fundamental origin of this good agreement between the experimental data and the simulation result, indicating that the dominant nonlinear interband dynamics can be effectively described by an off-resonantly driven two-level model, despite the continuum of interband transition energies in bulk InSb.

## 3. Outlook

In this article, we presented the fundamental pillars for understanding low-energy nonlinear quantum phenomena using THz-2DCS in the first half and introduced comprehensive research work reported so far using this technique. Since THz-2DCS is still at an early stage, its future development may diverge in various directions. Nevertheless, one promising direction is the combination of THz-2DCS with additional optical pulses, as in one-dimensional optical-pump THz-probe techniques (OPTP) [79-81]. By controlling the properties of the optical pulse, such as its polarization and/or power, one can modulate the characteristics of a target sample "fully optically," and numerous studies using OPTP have been published so far. Thus, the extensive knowledge and experimental expertise accumulated through OPTP studies can be readily united to extend THz-2DCS into a combined THz-2DCS+OPTP platform. By precisely controlling the optical pumping timing, this approach is expected to enable the investigation of low-energy nonlinear carrier dynamics as a function of parameters modulated by external optical pulses, which offers the advantage of not requiring repeated sample alignment. In addition, combining THz-2DCS with OPTP may help address the coherent-artifact problem in collinear THz-2DCS measurements originating from nonlinear responses of the EO detection crystal. By independently modulating the optical pump, the pump-dependent sample response can be isolated from the artifacts that are not modulated by the optical excitation.

## REFERENCEES

**Acknowledgements**

This research was supported by Global-Learning & Academic research institution for Master's · PhD students, and Postdocs(LAMP) Program of the National Research Foundation of Korea (NRF) grant funded by the Ministry of Education(No. RS-2024-00442775), and the regional Innovation System & Education(RISE) Glocal University 30 Program through the Gwangju RISE Center, funded by the Ministry of Education(MOE) and the Gwangju Metropolitan City, Republic of Korea.((2026-RISE(Glocal University 30)-05-011), and the ministry of education(moe) and the jeonnam-gwangju special metropolitan city, republic of korea (2026-anchor-05-011) (S. J. P and J. W. H).

**Author contributions:**

J.W.H. designed and supervised the review. S. J. P. performed simulations. S.J.P. and J.W.H. wrote the original manuscript. I.H.C. and J.W.H. revised the manuscript.

**Conflicts of Interest**

The authors declare no conflicts of interest.

**Data Availability Statement:** All data needed to evaluate the conclusions in the paper are present in the paper and/or the Supplementary Materials. Additional data related to this paper may be requested from the authors.